\documentclass[12pt,a4paper]{article}
\usepackage{amsmath,amssymb,amsthm,epsfig}

\newcommand{\bra}{\langle}
\newcommand{\ket}{\rangle}
\newcommand{\tr}{{\rm tr}}

\def\be#1\ee{\begin{equation}#1\end{equation}}
\def\bea#1\eea{\begin{align}#1\end{align}}

\title{On Quantum Theory}
\author{Rudolf Haag}
\date{}              % Activate to display a given date or no date

\begin{document}
\maketitle
%\bibliographystyle{plain}
%\bibliography{rh}

\abstract{\noindent 
A discussion of fundamental aspects of quantum theory is presented, 
stressing the essential role of ``events''.}\footnote{Abstract by Erhard 
Seiler (see afterword)}

\parskip=2mm

\section{Metaphysics}
%\subsection{}

Looking superficially at so many shelves with books and journals in a 
physics library we recognize already some basic features. 

First: Physics is a collective enterprise needing exchange between many 
human individuals. Each person experiences sensations registered in his 
conscious mind. And the remarkable thing is that we can communicate with 
others and find a consensus about some part of such impressions. The 
recognition of other persons with minds independent of ours together with 
the possibility of communicating with them suggests that we are small 
parts of a huge world which we call nature or the universe, most of which 
we assume to be disjoint from human consciousness. The complement of 
conscious minds we call the outside world.

Since everything we can ever know rests ultimately in the consciousness of 
our minds, there is the question stressed in idealistic philosophy as to 
whether the assumption of an outside world is superfluous. Pursuing this 
we may arrive at the idea of a universal consciousness constituting the 
universe. Such philosophical endeavors must be understood as the search 
for a metaphysical model appropriate to the motivation of our actions. In 
the fields of fine arts, music, religion, the emphasis on the primary role 
of consciousness is natural, it does, however, not provide the motivation 
for natural sciences, particularly not for experimental physics.

Here we take the dualistic or realistic point of view assuming two 
separate realms: the outside world on the one hand, and consciousness on 
the other. This was the driving motivation of experimental physicists. 
Between the outside world and consciousness there is an interaction in 
both directions. On the one hand, there is the action of the outside world 
on us, causing impressions. On the other hand, the action from us on the 
outside world, producing a change there subject to our will. This is 
essential for physics proper, involving the planning and setting up of 
experiments and recognizing that the senses of touch and vision could be 
replaced and refined by instruments called detectors. We shall call the 
response of such a detector a (coarse) event. An important feature in this 
context is the possibility of creating documents of such events, which may 
be studied much later by anybody interested.

Let us make it clear that we do not imply that one or the other of these
metaphysical models is true or false. We take recourse to the
``Philosophie des Als ob'' (Philosophy of ``as if'') by Hans Vaihinger.
(allowing any view as long as it is not contradicted...).

Second: ``divisibility''. Any gain of knowledge starts with (needs) the 
distinction of different things: different individuals of any kind. Our 
ability to distinguish different elements leads us to the concept of 
numbers, sets, from which the whole structure of mathematics ultimately is 
built. Moreover, the recurrent appearance of simple parts of impressions 
leads us to assign symbols (words) to them, resulting in the development 
of language and the ability to communicate with others about our 
impressions. In the quasi-static case (e.g. a tree stands before my 
window) the symbol tree stands for a common element recognizable in an 
almost continuous succession of impressions, thereby integrating all of 
them into a single concept. We instinctively associate with the symbol 
“tree” an element of the outside world, interacting with our consciousness 
and causing the impression. The assumed element of the outside world 
responsible for such an impression is called an object. Typically it may 
be a piece of matter registered by our sense of touch or a ray of light 
interacting with our retina. The motivation for considering an object as 
an element of the outside world is: we can control the appearance by 
opening or closing our eyes, approach it with different senses, we can 
discuss it with our neighbors. In addition it satisfies our deeply rooted 
urge for causality, the demand that the impression of the tree should have 
a cause.

Throughout centuries of physics and chemistry the classification of stable 
objects, the structure of matter, has been pursued with enormous success. 
As the description progressed towards finer and finer features, the 
visualization of objects became more and more fuzzy, so that the concept 
of objects no longer appears to provide a suitable basis in theory. Much 
closer to reality is the concept of an event, marking the result of an 
interaction process. In quantum physics, it marks the transition from a 
possibility to a fact. Its salient properties are irreversibility and an 
approximate localization in space and time, which provides the basis for 
our analysis of geometric relations in space-time. 

For each person the primary phenomena (registered events) form a 
well-ordered sequence in time. The sequence is steadily growing. The last 
phenomenon in this sequence is of a special character, it is the only one 
felt directly. It marks the subjective present, the earlier ones are 
guarded in the memory or in records, the later ones are not (yet) 
existing. The present divides time into past and future. The past consists 
of unchangeable facts, the future is open. 

Any statements about future events depend on inductive reasoning. 
Observations of regularities in the occurrence of past events have led to 
the formulation of “laws of nature”. We stipulate that these will hold in 
the future. Thus, they allow us to make predictions about coming 
phenomena.

In classical theory the laws are strictly deterministic, i.e. precise 
knowledge of objects and events in the past (unattainable in practice) 
would enable us to make precise predictions for the development in the 
future. Moreover it is widely believed that these laws satisfy complete 
symmetry between past and future (time reversal invariance); we shall 
return to this later. 

In quantum theory past history does not allow to predict the future; the 
past determines only probabilities of future events. Furthermore, the 
observation of an event does not allow us to make retrodictions of 
probabilities of past events (see Appendix A). This has some bearing on 
the question of reversibility, as discussed in the next section. 

\section{Irreversibility} 

The irreversibility of all naturally occurring processes is a general 
experience. This is most strikingly evident in biological processes (the 
development of organisms between birth and death is obviously 
unidirectional). But it is also omnipresent in physics and expressed there 
by the second law of thermodynamics. The reconciliation of this law with 
the generally assumed time reversal invariance of the basic laws of nature 
in classical physics is the challenge addressed by statistical mechanics. 
There, Boltzmann attributed to a macroscopic state a “thermodynamic 
probability” reflecting the number of different realizations of this 
macroscopic appearance by microscopic configurations. He interpreted this 
probability as the logarithm of the  ``entropy´´, whose growth with 
increasing time is stipulated in the second law of thermodynamics. These 
probabilities differ enormously between various macroscopic states and it 
is plausible to expect that under many circumstances the macroscopic body 
will change its state to one with higher thermodynamic probability in the 
passage of time. 

This is for instance so in laboratory physics, where the experimentalist 
sets the initial time at which a situation far from equilibrium (i.e. with 
low probability) is artificially created. It is also illustrated by  
dissipative effects, where typically a simple physical system describable 
by a few degrees of freedom begins to share energy with many untraceable 
degrees of freedom. Boltzmann’s argument by itself, however, does not 
suffice to justify the second law of thermodynamics. The differences in 
thermodynamic probability between various macroscopic states by themselves 
do not logically imply any distinction between past and future. We have to 
put in the additional information that at the beginning we start with a 
state of low probability. 

Irrespective of this, our above remarks about events in quantum physics 
point to a more elementary justification of irreversibility. We started 
with phenomena directly registered in our conscious mind and observed 
their sequential character leading to the “psychological” arrow of time. 
We then related these mental impressions to “events”. We shall show in the 
next section that the growth of the entropy is directly connected to the 
occurrence of events. Therefore, the thermodynamic arrow of time is 
coupled to the psychological arrow. We can adapt this argument to the case 
of classical physics. The notion of “event” must there be replaced by that 
of a process in which originally independent and distinguishable objects 
begin to interact, wiping out their distinguishability. 

We sometimes wonder why the dogma of time reversal invariance of the 
fundamental laws is so firmly entrenched, since it is not always clear 
what should be regarded as a fundamental law. In classical physics there 
is at least one aspect of dissipation that is of a fundamental nature. 
This is the radiation damping associated with any acceleration of a 
charged particle. In spite of Dirac’s beautiful analysis \cite{dirac} and 
many subsequent elaborations such as \cite{haag1955}) no formulation free 
from objections exists. Thus, strictly speaking, no complete 
self-consistent classical theory uniting mechanics and electrodynamics 
exists.

In Quantum Physics this is related to the infra-red problem and leads to 
the observation that the causal shadow of each event is its forward light 
cone. This leads to a formulation of the dynamics in terms of a semi-group 
shifting light cones inside each other, as worked out by Buchholz and 
Roberts \cite{Buchholz:2013bia}.

The strongest argument for the belief in time reversal invariance is the 
CPT theorem [footnote: T stands for Time reversal, C for Charge 
conjugation, P for space reflection], which states that in any quantum 
field theory satisfying the standard axioms there exists a CPT operator, 
implying that the equations of motion are CPT invariant. But, these 
equations of motion do not tell us what actually happens.  They only 
describe the development of the so-called “quantum state”, which 
determines the probability for the realization of different possible 
events. To arrive at facts (individual events) we must add to these 
equations the statistical interpretation of the quantum state, which in 
our context we might call the principle of random realization 
(\cite{haag2013}). 

It is this principle which destroys time reversal invariance and unites 
the various “arrows of time” discussed in the literature. To repeat the 
main line of our reasoning: we started with phenomena directly registered 
in our conscious mind and observed their sequential character leading to 
the “psychological” arrow of time. We then related these mental 
impressions to “events”, namely interaction processes which may be 
attributed to an outside world, and transported the sequential character 
from the phenomena to the events. Irreversibility happens in the 
transition from a possibility to a fact, which in quantum physics is 
governed by the principle of random realization, whereas in classical 
physics the state itself is supposed to describe reality and its 
deterministic propagation leaves no room for a distinction between 
possibilities and facts. 

\section{Standard Questions and Procedure}
\subsection{Observables} 

One decisive difference between the classical and quantum descriptions of 
nature is seen in the following statement of Niels Bohr: ``We cannot 
assign any conventional attributes to an atomic object''. As 
``conventional'' we may regard an attribute describable numerically. Of 
course there are defining properties of an object, such as mass and charge 
of an electron; but beyond this in classical physics we imagine that an 
object may appear in variable states, for instance that a particle has a 
definite position at any given time. Bohr's statement negates this.

Instead of attributes of an object one uses the concept of “observables”. 
This corresponds to the idea that one can carry through various 
measurements on an object, each of them leading to a numerical value, “the 
observational result”. The difference is that before the measurement no 
definite numerical value is assignable to the object. This value arises 
only by the interaction between the object and the measuring device. Every 
single observed value has to be considered a fact. As such it is 
irreversible, usually approximately localized in space and time; 
thus it corresponds to what we called an event above.

As a simple example of an object let us consider an electron. One possible 
observable is its position at a given time. It may be observed as a dot 
appearing at that time on a scintillation screen. This cannot be regarded 
as the position of the electron before the measurement, but is created by 
the interaction between the electron and the screen. The coordinates of 
this dot are not attributes of the electron but attributes of the 
interaction event.

A more interesting example is to choose as observable the internal energy 
of an atom. The possible measuring values are the energy levels of the 
electrons in the atom. These will manifest themselves by spectral lines 
corresponding to frequencies of the emitted light, which are proportional 
to energy differences. These frequencies in turn may be determined for 
instance by letting the light pass through a diffraction grating and 
recording the outcome on a photographic plate. Thus the observation of 
energy levels ultimately again leads to a position measurement.  We may 
note that in the last resort observations always terminate in a position 
measurement, since they record the location of an event.

The theoretical description uses the mathematics of Hilbert space. An 
observable is represented by a self-adjoint operator. These operators have 
a spectral resolution, i.e. spectral values and spectral projectors; the 
former are interpreted as possible measuring values.

The second characteristic feature of quantum physics is its lack of
determinism. It only yields statistical predictions for the outcome
of observations. These are governed by the so-called quantum state
which is mathematically represented by a positive operator $\rho$ of unit
trace, called the statistical operator. The probability of finding a 
measuring value in an interval $[a,b]$ is given by $\tr(\rho \ P_{[a,b]})$, 
where $P_{[a,b]}$ denotes the spectral projector for the interval [a,b]
and $\rho$ describes the quantum state before observation.

Since the theory yields only statistical predictions we need for checking 
them an ensemble of copies of equally prepared objects and we estimate 
probabilities by relative frequencies of occurrence. The realistic 
ensemble used in this test is idealized to a Gibbsian ensemble of 
infinitely many independent copies.

\subsection{Geometry of quantum states; composition and decomposition 
of systems; entanglement}

On the mathematical side the set of positive operators with unit trace is 
a convex set which means that for any two such operators $\rho_1,\rho_2$ 
all the convex combinations $\lambda \rho_1 + (1-\lambda) \rho_2$ with 
$0\le \lambda\le 1$ belong again to this set. This corresponds to the 
possibility of mixing the two ensembles with weights $\lambda$, 
$1-\lambda$, respectively.

The convex body of states has extremal points, the ``pure states'' which 
cannot be written as convex combinations of others. They are represented 
by one-dimensional projectors or equivalently by 
the rays in Hilbert space on which these project. In Quantum Mechanics of 
particles the wave function describes such a ray. The salient feature of 
quantum physics is that the convex body of states is not a simplex. Thus, 
while an arbitrary state can be written as a convex combination of pure 
states, this decomposition is highly nonunique.  In physical terms the 
decomposition of a state into a convex combination corresponds to a 
decomposition of the ensemble into subensembles.  Therefore it is often 
not possible or meaningful to assume that each individual system is in 
some pure state. The assignment of a particular pure state to an 
individual system means only that this system is filed as a member of a 
particular subensemble whose choice remains to some extent arbitrary.

This nonuniqueness has been the source of long disputes beginning with the 
EPR ``paradoxon'' \cite{EPR}, the concept of entanglement between states 
introduced by Schr\"odinger \cite{schr}, the inequalities by Bell 
\cite{bell} and Clauser et al \cite{clauser} and their subsequent 
experimental study by Aspect \cite{Aspect:1981nv, Aspect:1982fx} and many 
later experiments. 

For this discussion we must consider general systems composed of several 
subsystems. The Hilbert space associated with such a system is taken to 
be the tensor product of the Hilbert spaces of the subsystems which 
compose it. Its set of quantum states contains ``product states'' 
$\rho^{(1)} \otimes \rho^{(2)} \otimes ...$ and convex combinations 
thereof, called separable states, which describe correlations well known 
in classical statistics. These types of states do not, however, exhaust 
all possibilities. For instance, the pure states of the compound system 
are by definition not convex combinations of any other states, hence 
they cannot be separable unless they are just product states. States 
which are not separable are called ``entangled''. The statistical 
predictions for such entangled states cannot be described in terms of 
correlations between individual states of the subsystems. A simple 
illustration of this feature is afforded by a thought experiment 
suggested by Bohm \cite{bohm} and analysed in \cite{bell} and 
\cite{clauser}:

A spin-0 particle decays into two spin-1/2 particles moving in opposite 
directions for a long time till one of them enters the lab of Alice, the 
other one the lab of Bob. In both cases the arriving particle is subjected 
to a measurement of the spin orientation by a Stern-Gerlach arrangement. 
This can yield two possible outcomes: parallel or antiparallel to the 
orientation of the Stern-Gerlach magnet. We denote this result by $(\bf 
a,\alpha)$, where $\bf a$ is the unit vector describing the direction of 
the magnet; $\alpha=\pm1$ distinguishes the two possible results. The 
spin-part of the two-particle wave-function after the decay is a singlet 
state and this will remain so practically unchanged up to the detection 
process. This singlet state is a pure entangled two-particle state and one 
can show that it is impossible to assign any ``conventional attributes'' 
(``hidden variables'') nor even a quantum state to the individual 
particles. The former impossibility has been demonstrated in \cite{bell} , 
the latter in \cite{clauser} .We shall follow here the arguments of 
\cite{clauser}, as presented in \cite{speak}.

The ensemble of all particles received by Bob may be describerd by an 
impure one-particle quantum state $\rho_B$. Since the twin particles are 
correlated due to their common birth it is not surprising that the 
probability for a particular measuring results of Bob is correlated with the 
result of Alice's measuremant on the twin. However, entanglement is more 
than ordinary correlation.

Suppose now that a particle is endowed with some objective property 
$\lambda$ (which may be a quantum state or a conventional attribute) and 
the joint probability in the ensemble of pairs of particles is given by a 
distribution function $f(\lambda_1,\lambda_2)$ which describes ordinary 
correlation between $\lambda_1$ and $\lambda_2$. We assume further that 
$\lambda$ determines the probability $w(\lambda;\bf a,\alpha)$ for the 
measuring result $(\bf a, \alpha)$, yielding the expectation value 
conditioned on $\lambda$
\be
\bra {\bf a};\lambda \ket =w(\lambda;{\bf a},+)- w(\lambda;{\bf a},-)\,.
\ee  
We note that $w(\lambda;{\bf a},+)+w(\lambda;{\bf a},-)=1$ because in 
the measurement of $\bf a$, one of the alternatives $\pm 1$ must occur. 
The joint probability for $({\bf a}, \alpha;{\bf b}, \beta)$ is then 
\be
W({\bf a}, \alpha; {\bf b},\beta)=\int d\lambda_1 d\lambda_2 
f(\lambda_1,\lambda_2) w(\lambda_1;{\bf a},\alpha)
w(\lambda_2;{\bf b},\beta)\,.
\ee
For the expectation value in the joint measurement, which is defined by
\be
\bra{\bf a}; {\bf b}\ket\equiv W({\bf a}, +; {\bf b},+)-
W({\bf a}, +; {\bf b},-)- W({\bf a}, -; {\bf b},+)+
W({\bf a}, -; {\bf b},-)
\ee
one obtains the representation
\be
\bra{\bf a}; {\bf b}\ket= \int d\lambda_1 d\lambda_2 
f(\lambda_1,\lambda_2)
\bra {\bf a};\lambda_1 \ket \bra {\bf b};\lambda_2 \ket\,.
\ee
From this together with the positivity and normalization of the 
distribution function $f(\lambda_1,\lambda_2)$ one obtains inequalities 
between expectation values for combinations of measurements with different 
orientations of the apparatuses,
\be
|\bra{\bf a}; {\bf b}\ket+ \bra{\bf a}; {\bf b'}\ket|+
|\bra{\bf a'}; {\bf b}\ket-\bra{\bf a'}; {\bf b'}\ket|\le 2\,.
\ee
The experimentally observed violation of this inequality shows that the 
assumption of an ordinary correlation between assumed properties 
$\lambda_1,\lambda_2$ is not tenable. Instead one has the following 
situation: if Bob receives the information from Alice about what she has 
done and found in her measurements, he can split his ensemble into two 
subensembles according to Alice's measureing result $\alpha=\pm1$ on the 
twin. Then these subsensembles define two orthogonal pure states which 
depend on the orientation of Alice's device. It must be stressed that this 
has nothing to do with any physical effect of Alice's measurement on the 
particles received by Bob. Nor is it important how fast the information is 
transmitted. Bob and Alice can get together leisurely after the 
experiments are finished to evaluate their records. They only have to 
establish the correct pairing of the events, which can be found for 
example from the records of the arrival times. No witchcraft is involved.
It shows, however, that the pure state of the particle has no objective 
significance. It does not describe a property of the individual particle 
but only the defining information about the subensemble in which the 
particle is filed. And here this is determined by the result of Alices's 
measurement on the twin.

This implies an enhancement of Bohr's tenet mentioned in the introduction. 
Not only can we ``not assign any conventional attribute to an atomic 
object'' but we cannot even assign any individual quantum state to the 
particle. It puts in question the traditional picure of the reality of 
``atomic object'' (particles). Nicholas Maxwell has coined the term 
``Propensiton'' for such an object \cite{maxwell}. It propagates according 
to a deterministic law such as the Schr\"odinger equation which is 
invariant under time reversal. But it does not represent any phenomenon. 
It is the carrier of propensity contributing to probability assignments.

\appendix
\section{Temporal asymmetry of quantum probabilities}

For simplicity we consider here only two `observables' $A,B$ given by
self-adjoint operators with simple discrete spectrum. Assume that in a
state $\rho$ first A, then B is measured. Let $p_\alpha$ be the
probability that the measurement of $A$ yields the spectral value
$\alpha$, $p_{\alpha\beta}$ the probability that the consecutive
measurement of $B$ then yields $\beta$. We have, according to the L\"uders
rule \cite{luders}
\begin{equation}
p_\alpha= \tr(P_\alpha \rho),\quad p_{\alpha\beta}=
\tr (Q_\beta P_\alpha\, \rho\, P_\alpha Q_\beta)\,.
\label{pab} 
\end{equation}
Using the assumptions about the spectra we have
\begin{equation}
P_\alpha \rho P_\alpha = P_\alpha \tr(\rho P_\alpha)
\end{equation}
and hence, using the cyclicity of the trace
\begin{equation}
p_{\alpha\beta}= \tr (P_\alpha Q_\beta) \tr(P_\alpha\rho)
=|\bra\alpha|\beta\ket|^2 \bra
\alpha|\rho|\alpha\ket=p_\alpha|\bra\alpha|\beta\ket|^2 \,.
\end{equation}
The last formula shows already the asymmetry between the two measurements. 

More explicitly, when $\alpha$ has been measured, the probability of 
finding $\beta$ afterwards is known and given by that formula; if, on the 
other hand, we only know that $\beta$ has been measured in the second 
step, the probability that the first measurement yielded $\alpha$ cannot 
be inferred unless $|\alpha\ket=|\beta\ket$ (in which case it is 1).

\section*{Afterword\footnote{by Erhard Seiler (Max-Planck-Institut f\"ur 
Physik, M\"unchen, Germany, e-mail: \\
\hglue20pt  ehs@mpp.mpg.de)} }

Rudolf Haag worked on this manuscript since the spring of 2013 until just
a few weeks before his death on January 5th, 2016. It evolved in numerous 
discussions with me as well as  Heide Narnhofer and Berge Englert. 
Technical help was also provided by Albert Haag and Friedrich Haag.  But 
I should stress that the thoughts and the wording are entirely due to 
Rudolf Haag. 
\vskip10pt
Obviously the paper has not been completed; among the further issues 
that Rudolf wanted to address were

\begin{itemize}
\item
Partition of the universe, classification and reality of objects,
reductionism and its possible limits;
\item
Indistinguishability of particles; distinguishability of particles as
connections between events;
\item
Space and time, symmetries and invariances.
\end{itemize}

\end{document}